\documentclass[12pt]{article}
\pdfoutput=1

\usepackage{amssymb,amsmath,amsthm}
\usepackage{graphicx}
\usepackage[section]{placeins}
\usepackage[authoryear]{natbib}	
\usepackage[english]{babel}
\usepackage{url}
\usepackage{enumitem}
\usepackage{color}
\usepackage{dsfont}
\usepackage{subcaption}
\usepackage{multirow}
\usepackage{authblk}
\usepackage[hidelinks]{hyperref}

\def\spacingset#1{\renewcommand{\baselinestretch}%
	{#1}\small\normalsize} \spacingset{1}

\date{}

\begin{document}

\title{\textbf{Mixture Models and Networks - Overview of Stochastic Blockmodelling}}
\author{Giacomo De Nicola, Benjamin Sischka and G\"{o}ran Kauermann}

\affil{Department of Statistics, Ludwig-Maximilians-Universit\"{a}t M\"{u}nchen}

\maketitle

\begin{abstract}
	Mixture models are probabilistic models aimed at uncovering and representing latent subgroups within a population. In the realm of network data analysis, the latent subgroups of nodes are typically identified by their connectivity behaviour, with nodes behaving similarly belonging to the same community. In this context, mixture modelling is pursued through stochastic blockmodelling. We consider stochastic blockmodels and some of their variants and extensions from a mixture modelling perspective. We also survey some of the main classes of estimation methods available, and propose an alternative approach. In addition to the discussion of inferential properties and estimating procedures, we focus on the application of the models to several real-world network datasets, showcasing the advantages and pitfalls of different approaches.
\end{abstract}

\noindent

\textbf{Keywords:} {\it Community Detection; Mixture Models; Network Analysis; Stochastic Blockmodels.} 

\vfill

\newpage
\section{Introduction}
\label{sec:intro}

The underlying idea of a mixture model is rather simple.
Instead of assuming that the target variable follows a plain distribution, one considers a mixture of multiple distributions.
Specifically, for a random variable $Y$, one assumes 
\begin{align}
Y \sim \sum_{k=1}^{K} \pi_k f_k(y),
\label{eq:finmix}
\end{align}
where $\pi_k$ is a weighting coefficient, with $\sum_{k=1}^K \pi_k = 1$, and $f_k(\cdot)$ is the $k$-th mixture distribution. Commonly, the mixture components come from the same distributional family but differ in their parameters, that is 
$f_k(\cdot) = f(\cdot|\boldsymbol{\theta}_k)$
where $\boldsymbol{\theta}_k$ parametrizes the $k^{th}$ mixture component. 
An early (maybe the first) reference in this direction dates back to \cite{Pearson:94} and focuses on the estimation of a mixture of two normal distributions. 
An early mathematical treatment of the topic, more in the style of convolution, is provided in \cite{Robbins:48}.
In a series of papers, \cite{Teicher:60} discusses identifiability issues, where the cited work puts the focus on finite mixtures in the style of (\ref{eq:finmix}).
A first survey on mixture models is provided by \cite{GuptaHuany:81}, presenting the different estimation routines that had been developed and used by that time. 
A central algorithm in this respect, which is not included in the above survey article (certainly because of simultaneous time of publication), is the work of \cite{AitkinWilson:80} (see also \citealp{Aitkin:80}) who propose the use of the at the time recently developed EM algorithm (see \citealp{DempLairdRubin:77}) to estimate the finite mixture distribution. Though the focus of their paper lies in the modelling of outliers, the authors make use of the idea that a finite mixture model can be comprehended as a missing data problem. 
Under this modelling framework, one assumes that the discrete valued random variable $Z$ takes values $\{1,...,K\}$ with 
\begin{align}
P(Z=k) = \pi_k
\label{eq:Z}
\end{align}
for $\sum_{k=1}^K \pi_k = 1$. 
Conditional on $Z=k$ one then observes $Y$ from the $k$-th mixture component, i.e.\
\begin{align*}
	Y|(Z=k) \sim f_k(y) \quad \mbox{for } k=1,...,K.
\end{align*}
Treating $Z$ as unobserved (or unobservable) enables the framing of estimation in a missing data situation, where the considered likelihood (\ref{eq:finmix}) can be maximized with the EM algorithm. The results are generalized and extended towards hypothesis tests in \cite{AitkinRubin:85}. 
A comprehensive overview on finite mixture models is given in the early book of \cite{EverittHand:80}, followed by the monographs of
 \cite{TitteringtonSmithMavov:85}, \cite{Lindsay:95}, \cite{Boehning:99}, \cite{MacLachlanPeel:00}, or \cite{Fruehwirth:06}. We also refer to the recent Handbook of Mixture Analysis (\citealp{Fruehwirth:19}).
For software implementations of mixture models, \cite{Leisch:04} is a central reference (see also \citealp{Benagalia:09}). 
Allowing the mixture components and/or the mixing proportions $\pi_k$ to depend on additional covariates extends mixture models towards regression models. The resulting model class is also known as mixture of experts, tracing back to \cite{Jacobs:91}. 
A survey from the perspective of Machine Learning can be found in \cite{MasoudniaEbra:14}.
See also \cite{GormFrueh:19} for more details.

While most of the literature cited above deals with a univariate response variable $Y$, in this paper we aim to look at multivariate data with $\boldsymbol{Y}$ expressing a network.
Network data have a simple binary structure resulting from a network as follows. Assume a set of $N$ actors, where we define with $V = \{v_1,...,v_N\}$ the set of nodes in a network. We call $E\subset V\times V$ the edge set, and the resulting network can be represented with an adjacency matrix $\boldsymbol{Y}$ such that $\boldsymbol{Y} \in \{0,1\}^{N\times N}$ and 
$$ Y_{ij} = \begin{cases}
1 \quad \mbox{if } (v_i, v_j) \in E \\
0 \quad \mbox{otherwise.}
\end{cases} $$
If the network is undirected, $Y_{ij} = Y_{ji}$ holds. Furthermore, the diagonal of $\boldsymbol{Y}$ often remains undefined, meaning that self-loops are not contemplated. The statistical analysis of network data has achieved increasing interest in the last decade: we refer to \cite{Kolaczyk:09} and \cite{KolaczykCsardi:14} for a general introduction to the topic (see also \citealp{Goldenberg:10}, \citealp{Hunter:12}, \citealp{Fienberg:12}, \citealp{Lusker:13}, or \citealp{BiageniKauermannMeye:19}). 

If we consider $\boldsymbol{Y}$ as set of random variables $\{Y_{ij}, 1 \leq i,j \leq N, i \neq j \}$, we can transfer the mixture model setting (\ref{eq:finmix}) towards network data. This leads to so-called (\textsl{{a posteriori}}) stochastic blockmodelling, which can be seen as a tool for performing \textsl{community detection}. A survey on the latest theoretical developments in this field has recently been published by \cite{Abbe:19}. While  community detection and stochastic blockmodels have a lot in common, the latter specifically focusses on the modelling aspect and will therefore be considered here. A stochastic blockmodel (SBM) is in fact a mixture model where each mixture component is specified by the group or community membership. Hence, we assume the independent discrete group indicator coefficients $Z_i \in \{1,...,K\}$ for $i=1,...,N$ with
$$ \mathbb{P}(Z_i = k) = \pi_k \quad \mbox{for } k=1,...,K $$
and, as above, $\sum_{k=1}^K \pi_k=1 $. An edge between node $i$ and $j$ then exists with probability 
\begin{align}
Y_{ij}|\boldsymbol{Z} = \boldsymbol{z} \overset{ }{\sim} Bernoulli(p_{z_{i} z_{j}}),
\label{eq:sbm}
\end{align}
where $\boldsymbol{P} = [p_{kl}]_{k,l=1,\ldots, K}$ is the ${K \times K}$ dimensional block-probability matrix. 
For community detection one typically assumes that $p_{kk} > p_{kl}$ for $l \neq k$, but this is not a requirement for stochastic blockmodels in general. In fact, the block structure may describe clusters of nodes that behave similarly from a connectivity standpoint without necessarily being more densely connected, thus allowing for other types of structures, such as disassortative communities and core-periphery.

The class of stochastic blockmodels evolved from its deterministic counterpart, which dates back to \cite{White}. The stochastic version of the blockmodel was introduced by \cite{HollandLaskeyLein:83} in the statistical literature. Similar modelling proposals, developed independently, trace back to the computer science literature, see e.g.\ \cite{Bui:87}. \cite{Wang1987} were the first to apply the stochastic blockmodel to directed graphs, even though they still assumed the block structure to be known. The first steps towards so-called \textsl{a posteriori} blockmodelling, that is, modelling with initially unknown group structure, were taken by \cite{Snijders1997} and \cite{Nowicki2001}, who proposed estimation routines for, respectively, two groups and any known number of groups. From there, the model class gained traction. Recent literature on the classical version of stochastic blockmodels include \cite{GormleyMurphy:10} or \cite{Aitkin:14}, using Bayesian approaches (see also \citealp{Vu2015}).

Stochastic blockmodels have been extended in various ways, some of which we will discuss in this paper. Well known is the \textsl{degree-corrected} stochastic blockmodel, introduced by \cite{Karrer2011}. In their work, the authors show how the standard stochastic blockmodel implicitly assumes the degree structure within communities to be relatively homogeneous. This, combined with the fact that many real world networks exhibit extremely skewed degree distributions \citep{Simon1955, Barabasi1999}, leads the model to often only be able to find core-periphery type block structures, with nodes grouped mostly on the basis of degree similarity. To bypass this issue, \cite{Karrer2011} introduced the idea of degree correction, making the probability of an edge depend not only on group membership, but also on node-specific heterogeneity parameters. 
Other notable extensions of the stochastic blockmodel include the mixed membership model \citep{Airoldi2008}, in which nodes can belong to multiple communities simultaneously, and the hierarchical stochastic blockmodel \citep{Peixoto2017}, in which communities are comprised of meta-communities, leading to a hierarchical block-structure. It is also possible to add covariates to the analysis, as initially proposed by \cite{Tallberg2005}. All of the mentioned specifications can be applied to binary data as well as to valued and count data (see e.g.\ \citealp{Nowicki2001}). In this paper, we do not concentrate on these extensions, but focus on more ``classical" SBMs. 

The rest of the paper is organized as follows: Section~\ref{sec:2} describes the blockmodelling framework in more detail. Section~\ref{sec:3} presents some real-world network datasets together with the potential questions that we face in the analysis of the networks. Section~\ref{sec:4} compares the different estimation routines that are available to answer the questions posed in Section~\ref{sec:3}, as well as proposing an alternative  method for the estimation of stochastic blockmodels. The results from the analyses are contained in Section~\ref{sec:5}. Finally, Section~\ref{sec:disc} ends the paper with some comments and conclusive remarks.

\section{Variants of the Stochastic Blockmodel}
\label{sec:2}

The ``original'' stochastic blockmodel is given in (\ref{eq:sbm}).
For estimation, a numerically simpler setting results by approximating the binomial distribution through a Poisson distribution. This approximation is justified since the network density is usually low, implying that $p_{kl}$ is typically small. In this case, (\ref{eq:sbm}) is replaced by 
\begin{align}
Y_{ij}|(\boldsymbol{Z} = \boldsymbol{z}) \overset{ }{\sim} Poisson(\lambda_{ij})
\label{eq:4}
\end{align}
where $\lambda_{ij}=\exp\{{\omega_{z_{i} z_{j}}}\}$ with $\boldsymbol{\Omega} = [ \omega_{kl}]_{k,l=1,\ldots, K}$ as block-connectivity para\-meter matrix. 


A further extension of (\ref{eq:4}) and hence (\ref{eq:sbm}) results through so-called degree-corrected stochastic blockmodels, which allow for node-specific heterogeneity effects. More precisely, the original version of the degree-corrected SBM can be written in the same way as (\ref{eq:4}), but in this case
\begin{align}
 \lambda_{ij}=\exp\{{\gamma_i+\gamma_j+\omega_{z_{i} z_{j}}}\}.
 \label{eq:degcor}
\end{align}
In this notation $\exp\{\gamma_i\}$ quantifies the heterogeneity specific of node $i$, and $\exp\{\omega_{z_{i} z_{j}}\}$ can be viewed as a measure of the propensity to form ties between the groups to which nodes $i$ and $j$ belong. All three versions, namely (\ref{eq:sbm}), (\ref{eq:4}), and (\ref{eq:degcor}), will be applied to data examples introduced in the next section. 


\section{Data Description}
\label{sec:3}

In order to demonstrate the capabilities of stochastic blockmodels we have chosen network datasets pertaining to three different domains, namely political science, biology and sociology. Despite the different domains, the networks share the presence of some form of underlying community structure, or at least the appearance thereof. They all therefore lend themselves to be modeled through the use of mixture components. General descriptive measures of the examples are given in Table \ref{tab1}, which shows that all three networks are of medium size and range from very dense to relatively sparse. 
\begin{table}[tbh]
	\center
	\begin{tabular}{l|llll}
		& Alliances & Butterflies & E-Mails &  \\\hline
		Nodes & 141 & 832 & 548 &  \\
		Edges & 1703 & 86528 & 5433 &  \\
		Density& 0.173 & 0.250 & 0.036 & \\
	\end{tabular}
	\caption{Descriptive statistics for the studied networks.}
	\label{tab1}
\end{table}

\subsection{International Alliances Network}

The first network that we introduce is constructed using data from the Alliance Treaty Obligations and Provisions project \citep{Leeds2002}. The dataset provides information on military alliance agreements pertaining to all countries of the world. For the analysis we consider alliances that were in force in the year 2016. The countries are taken as nodes, and an edge between two countries is present if the two countries take part in a ``strong'' military alliance treaty. More specifically, the alliances that we consider strong are {defensive} and {offensive} ones. This means, respectively, ``\textsl{alliances in which the members promise to provide active military support in the event of attack on the sovereignty or territorial integrity of one or more alliance partners}'' and ``\textsl{alliances in which the members promise to provide active military support under any conditions not precipitated by attack on the sovereignty or territorial integrity of an alliance partner, regardless of whether the goals of the action are to maintain the status quo}'' (see \citealp{Leeds2002}). 

Looking at this network from a blockmodelling perspective, there are several questions that we can pose. First of all, do the alliances between countries induce a partition of the network that is meaningful from a geopolitical perspective? Moreover, will the blocks found be in line with geographic proximity and political affinity, or will there be some other characteristics driving the grouping? And finally, what can the resulting block-structure tell us about the global system of alliances?


\subsection{Butterfly Similarity Network}

The second real-world instance is a butterfly similarity weighted network, constructed using the data presented by \cite{Wang2009} and available from \cite{biosnapnets}. Nodes represent butterflies and valued edges depict visual similarities between them. The similarity scores lie in the interval $[0,1]$, with a higher value implying a higher level of similarity. Scores
are computed using butterfly images, as described in \cite{Wang2009}. Information on the species to which each butterfly belongs is also available, with each unit belonging to a single species. A total of 10 species are present, implying a ``natural'' partition of the network in 10 blocks.

In this case, there is one clear question that emerges: are the communities found by using visual similarity scores in agreement with how biologists categorized butterfly species? In other words, are we able to recover the ``ground truth'' communities of the network via stochastic blockmodelling?

\subsection{Email Exchanges Network}

The last network consists of anonymized email data from a large European research institution collected between October 2003 and  May 2005 \citep{snapnets}. Each node in the network represents a person, and an edge between nodes $i$ and $j$ is present if person $i$ sent person $j$ at least one email in the examined period. The nodes featured in this network are all members of the institution, meaning that only emails within the institution itself are considered. Moreover, only nodes belonging to the largest ten departments are included.
Since department memberships are known and individuals from the same department are expected to behave similarly, we can consider the departments as ``ground truth'' communities for the network. Given that, the questions that we pose are straightforward: are we able to find some form of meaningful community structure in the network considering emails alone? 
And if so, will the structure recovered be similar to the partition induced by department memberships? And finally, what can email exchanges tell us about the structure of the institution and the relationships between departments?

Before analysing this and the other previously introduced networks and examining the correspondingly raised questions, we introduce the estimation procedures which are used to fit the appropriate model variants.

\section{Estimation Techniques}
\label{sec:4}

\subsection{Variational Methods}

The EM algorithm proved to be a powerful and numerically efficient way for estimating parameters in mixture models (see \citealp{Aitkin:80} or \citealp{Friedl2000}). Unfortunately, this does not extend to the estimation of stochastic blockmodels. The complete data log-likelihood resulting from (\ref{eq:4}) in the case of an undirected network equals
\begin{align}
l_C(\boldsymbol{\Omega},\boldsymbol{\pi})= \sum_{i=1}^N \sum_{j=i}^N \sum_{k,l=1}^K  \mathds{1}_{\{z_i=k\}}  \mathds{1}_{\{z_j=l\}}(y_{ij}\omega_{kl} - \exp\{{\omega_{kl}}\}) + \sum_{i=1}^N \sum_{k=1}^K \mathds{1}_{\{z_i=k\}}\log{\pi_k}
\label{eq:complete}
\end{align}
with the side constraint $\sum_{k=1}^K \pi_k = 1$. Applying the EM algorithm would in this case mean calculating the posterior distribution
$$ {P}(Z_i=k, Z_j=l \,|\, \boldsymbol{Y}= \boldsymbol{y}) $$
with $\boldsymbol{y}$ being the observed adjacency matrix. This posterior, due to the resulting dependence structure of $Z_i$ and $Z_j$, is numerically intractable \citep{Mariadassou2010}. To circumvent such numerical hurdles, \cite{Jordan1999} proposed variational methods, which are based on an approximation of the likelihood. Let ${P}(\boldsymbol{y};\boldsymbol{\Omega},\boldsymbol{\pi})$ be the probability of the data, resulting through
$${P}(\boldsymbol{y};\boldsymbol{\Omega},\boldsymbol{\pi}) = \sum_{k_1=1}^K \ldots \sum_{k_N=1}^K \pi_{k_1}...\pi_{k_N} \prod_{i=1}^{N} \prod_{j>i}^{N} \lambda^{y_{ij}}_{k_{i}k_{j}}\exp\{{-{\lambda_{k_{i}k_{j}}}}\}$$
which is apparently too complex from a numerical perspective. We define the lower bound function
$$J(\tilde{P}(\boldsymbol{z};\boldsymbol{\xi});\boldsymbol{\Omega},\boldsymbol{\pi})=\log P(\boldsymbol{y};\boldsymbol{\Omega},\boldsymbol{\pi})-\text{KL}(\tilde{P}(\boldsymbol{z};\boldsymbol{\xi}),P(\boldsymbol{z} \, |\, \boldsymbol{y};\boldsymbol{\Omega},\boldsymbol{\pi}))$$
where $\text{KL}(\,,\,)$ defines the Kullback-Leibler divergence. If we choose $\tilde{P}(\boldsymbol{z},\boldsymbol{\xi})$ to be the posterior distribution of $\boldsymbol{Z}$ given $\boldsymbol{\xi}$, we obtain $J(\, ; \,)$ to be equal to the log-likelihood of the observed data. Since this is numerically problematic, we compute the posterior distribution of $\boldsymbol{Z}$ given $\boldsymbol{\xi}$ through independence:
$$\tilde{P}(\boldsymbol{z};\xi)=\prod_{i=1}^{N} \prod_{k=1}^{K} {{\xi_k}^{\mathds{1}_{\{z_i=k\}}}}$$
where $\sum_{k=1}^{K}\xi_k = 1$. The parameter vector $\boldsymbol{\xi} = (\xi_1, \ldots , \xi_K)$  is known as variational parameter, and needs to be chosen such that $J(\tilde{P}(\boldsymbol{z};\boldsymbol{\xi});\boldsymbol{\Omega},\boldsymbol{\pi})$ is maximized with respect to all parameters. It can be shown that $J(\, ; \,)$ can, up to an intractable constant, be written in a simple numerical form which allows for fast and numerically feasible estimation. The remaining unknown component expresses the approximation error which is typically difficult to quantify (see \citealp{Lee2020}).

\subsection{Vertex Switching Algorithms}
\label{subSec:vsa}

Another possibility for the estimation of stochastic blockmodels is to maximize the likelihood through vertex switching routines. The basic idea of this type of local heuristic algorithms is the following: starting from an initial, possibly random group assignment, a starting value of the likelihood is computed. From there, one or more vertices are moved from one group to another, and the likelihood is computed again. The new allocation is then accepted or rejected based on a function of the two likelihoods, and such procedure runs iteratively until a maximum is found meaning until convergence is reached. 
Algorithms of this type include single-vertex Monte Carlo and a local heuristic routine inspired by the Kernighan–Lin algorithm used in minimum-cut graph partitioning \citep{Kernighan1970,Karrer2011}.
In principle, computing the likelihood that many times may seem quite expensive. On the other hand, it is not always necessary to calculate the complete likelihood at each step. Depending on the model specification, it is often possible to write the change in the likelihood in a computationally efficient way, so that the algorithm becomes quite competitive in terms of speed. 

The chief issue with this type of algorithm is that, given the heuristic maximization routine, it is not possible to obtain a measure of uncertainty for group assignments. The procedure will only produce the graph partitioning that maximizes the likelihood, without any additional information. This is fine if the problem at hand is one of pure community detection, but can become problematic if the goal is proper mixture modelling, as the stochastic component of the mixture is lost. Another potential issue is the possibility to get stuck at local maxima, which usually is tackled by running the procedure several times with different (random) starting points.

\subsection{Monte-Carlo-based EM Estimation}
A third somewhat novel estimation routine is to estimate the block structure using an EM-type algorithm including Gibbs sampling in the E step. As mentioned above, EM-based algorithms in mixture models are generally a natural choice, although numerically demanding. We therefore make use of approximations based on MCMC simulations. 
To do so, we slightly reformulate the SBM, relating it to graphon estimation (see e.g. \citealp{latouche2016} or, for a reverse link, \citealp{WolfeOl:14} and \citealp{Airoldi:13}). We here want to follow the estimation approach of \cite{kauermann2019}, applying it to SBMs. Regarding the reformulation, we introduce $U_i$ with $i=1,\ldots,N$ as continuous random variables within $[0,1]$ and divide the interval $[0,1]$ into $K$ sub-intervals, with $0=\tau_0 <\tau_1 < \ldots < \tau_K =1$ as boundaries. Model (\ref{eq:sbm}) is then rewritten to
\begin{gather*}
	U_i \overset{ }{\sim} Uniform[0,1] \quad i.i.d. \\
	Y_{ij} | \boldsymbol{U} = \boldsymbol{u} \overset{ }{\sim} Bernoulli (p (u_i,u_j)),
\end{gather*}
where $p: [0,1] \times [0,1] \rightarrow [0,1]$ is a function (sometimes called graphon) which is assumed to be local constant in the rectangles defined by the $K$ groups, that is 
\begin{align}
	p(u_i,u_j) = \sum_{k=1}^{K} \sum_{l=1}^{K} \mathds{1}_{\{\tau_{k-1} \leq u_i < \tau_k \}} \mathds{1}_{\{\tau_{l-1} \leq u_j < \tau_l \}} p_{kl}
	\label{eq:graphon}
\end{align}
with $p_{kl}$ as defined above for $k,l = 1 ,\ldots,K$. An instance of such relationship can be given through the following illustration:
\begin{center}
	\begin{minipage}[c]{0.48\textwidth}
		\begin{align*}
		\left.
		\begin{gathered}
		\boldsymbol{\pi} = (0.5,0.2,0.3) \\[0.2cm]
		\boldsymbol{P} = \begin{pmatrix}
		0.6 & 0.1 & 0.3\\
		0.1 & 0.5 & 0.2\\
		0.3 & 0.2 & 0.4
		\end{pmatrix}
		\end{gathered} \quad \right\} \quad \; \; \; \Longleftrightarrow
		\end{align*}
	\end{minipage}
	\begin{minipage}[c]{0.48\textwidth}
		\vspace*{0.1cm}
		\includegraphics[trim={1cm 1cm 0 0.5cm},clip,width=0.9\textwidth]{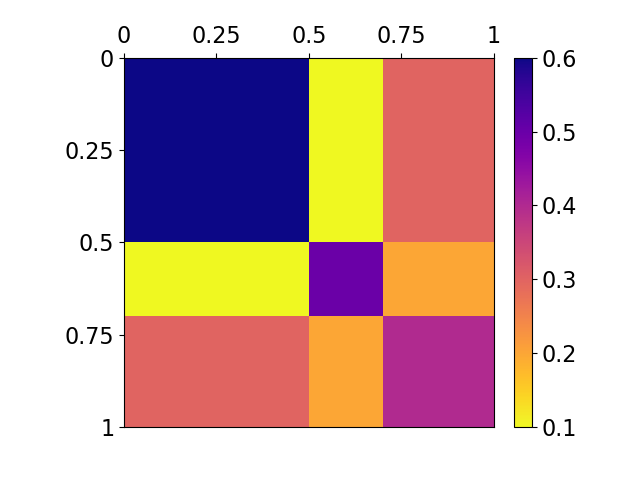}
	\end{minipage}
\end{center}
\vspace*{0.2cm}
It is not difficult to see how this model formulation is equivalent to SBMs. As such, we obtain the group probabilities through
\begin{align*}
	\pi_k = \tau_k - \tau_{k-1}.
\end{align*}
The idea is now to make use of model (\ref{eq:graphon}) to estimate both $p_{kl}$ as well as the interval boundaries $\tau_1, \ldots, \tau_{K-1}$. Assume the matrix $\boldsymbol{P} = [p_{kl}]_{k,l=1,\ldots, K}$ to be given (or set to a current value in the algorithm). Then, from (\ref{eq:graphon}), the full conditional posterior can be formulated as
$$
	g_j (u_j | u_1, \ldots, u_{j-1} , u_{j+1} , \ldots, u_N, \boldsymbol{y}) \propto \prod_{i=1}^N \prod_{j > i}^N p(u_i, u_j)^{y_{ij}} (1 - p(u_i, u_j))^{1-y_{ij}}.
$$
This allows for MCMC Gibbs sampling in a straightforward manner. We give details in the Appendix. We define the posterior mode in the $m$-th iteration with $\hat{u}_j^{(m)}$ for $j=1,\ldots,N$, where $\hat{u}_j^{(m)} = \tau_{k^{\prime}-1}^{(m)} + (\tau_{k^{\prime}}^{(m)} - \tau_{k^{\prime}-1}^{(m)})/2$ with index $k^{\prime}$ defined as $\mbox{arg max}_k \sum_t \mathds{1}_{\{\tau_{k-1}^{(m)} <  u_m^{<t\cdot r>} \le \tau_k^{(m)}\}}$.

The M step then is carried out by maximizing the likelihood, which is easily done by setting 
\begin{align*}
	\hat{p}_{kl}^{(m+1)} = \frac{\sum_{i=1}^{N} \sum_{j=1}^{N} \mathds{1}_{\{ \tau_{k-1} \leq \hat{u}_i^{(m)} < \tau_k \} } \mathds{1}_{\{ \tau_{l-1} \leq \hat{u}_j^{(m)} < \tau_l \} } y_{ij} }{\sum_{i=1}^{N} \sum_{j=1}^{N} \mathds{1}_{\{ \tau_{k-1} \leq \hat{u}_i^{(m)} < \tau_k \} } \mathds{1}_{\{ \tau_{l-1} \leq \hat{u}_j^{(m)} < \tau_l \} }}.
\end{align*}
It remains to update $\tau_k$, or equivalently, $\pi_k$. To do so, we set
\begin{align*}
	\hat{\pi}_k^{(m+1)} = \delta^{(m+1)} \frac{\sum_{i=1}^{N} \mathds{1}_{\{ \tau_{k-1} \leq \hat{u}_i^{(m)} < \tau_k \} } }{N} + (1 - \delta^{(m+1)}) \frac{1}{K},  
\end{align*}
where $\delta^{(m+1)}$ induces a step-size adaptation with $\delta^{(m+1)} \in [0,1]$ 
and $\delta^{(m+1)}>\delta^{(m)}$. Such step-size adaptation is recommendable
to prevent the community size to shrink too substantially before the structure of the community has been evolved properly. In general, $\delta^{(m+1)}$ is chosen to be one in the last iteration. 
As it is done for the vertex switching algorithms, we run this MCEM algorithm several times with varying initial values and then choose the outcome with the highest likelihood, which should here also prevent getting stuck at a local maximum. The advantage of the reformulation of model (\ref{eq:sbm}) or (\ref{eq:4}) to model (\ref{eq:graphon}) is that the graphon function $p(\, , \,)$ could also be made more complex, i.e.\ instead of just being local constant one could allow for more complex structures within each interval. This is not further discussed in this paper, but we refer to this new research strand discussed e.g.\ in \cite{vu2013}.  

In contrast to deterministic estimation routines, such as the vertex switching algorithm discussed in Subsection~\ref{subSec:vsa}, this modelling approach naturally yields information about the inherent uncertainty of the proposed group allocation. In order to achieve this, we run the E step one more time after the algorithm has converged. The resulting Gibbs sampling sequence of this last iteration then reveals the distribution of the node allocation with respect to the model estimate $(\hat{p}(\, , \,),\hat{\boldsymbol{\tau}} = (0,\hat{\tau}_1, \ldots, \hat{\tau}_{K-1},1))$. 
A normalised Gini coefficient calculated over the assignment frequencies of a single vertex can then be used as a measure of uncertainty, where a value near one (zero) implies a low (high) level of uncertainty.


\subsection{Choosing the Number of Blocks}

A general big challenge in mixture models (and hence also in stochastic blockmodels) 
lies in the choice of the number of mixture components (blocks). In fact, all the variants presented so far require that number to be known \textit{a priori}. This is typically not true in real-world applications. In mixture models the question of choosing the number of mixture components is tackled, for instance, in \cite{Aitkin2011}. 
In the field of stochastic blockmodels, approaches based on penalized likelihood criteria have emerged. In particular, \cite{Wang2017} consider an approach based on the log-likelihood ratio statistic, enabling the use of a likelihood-based model selection criterion that is asymptotically consistent. Other techniques are also available: \cite{Chen2018} develop a network cross-validation approach which is based on a block-wise node-pair splitting technique, combined with an integrated step of community recovery using sub-blocks of the adjacency matrix. \cite{Mariadassou2010} base the choice on an Integrated Classification Likelihood (ICL) criterion. Finally, \cite{Riolo2017} present a method for estimating the number of communities in a network using a combination of Bayesian inference and an efficient Monte Carlo sampling scheme.
While other approaches have been proposed, we will not go into further detail here. For modelling the previously described networks, we select $K$ such that the resulting number of blocks is reasonable and allows for appropriate interpretation.

\section{Applications}
\label{sec:5}

\subsection{International Alliances Network}

\begin{figure}[tbh]
	\centering
		\includegraphics[width=\linewidth,trim=0cm 9.5cm 0cm 9.5cm,clip]{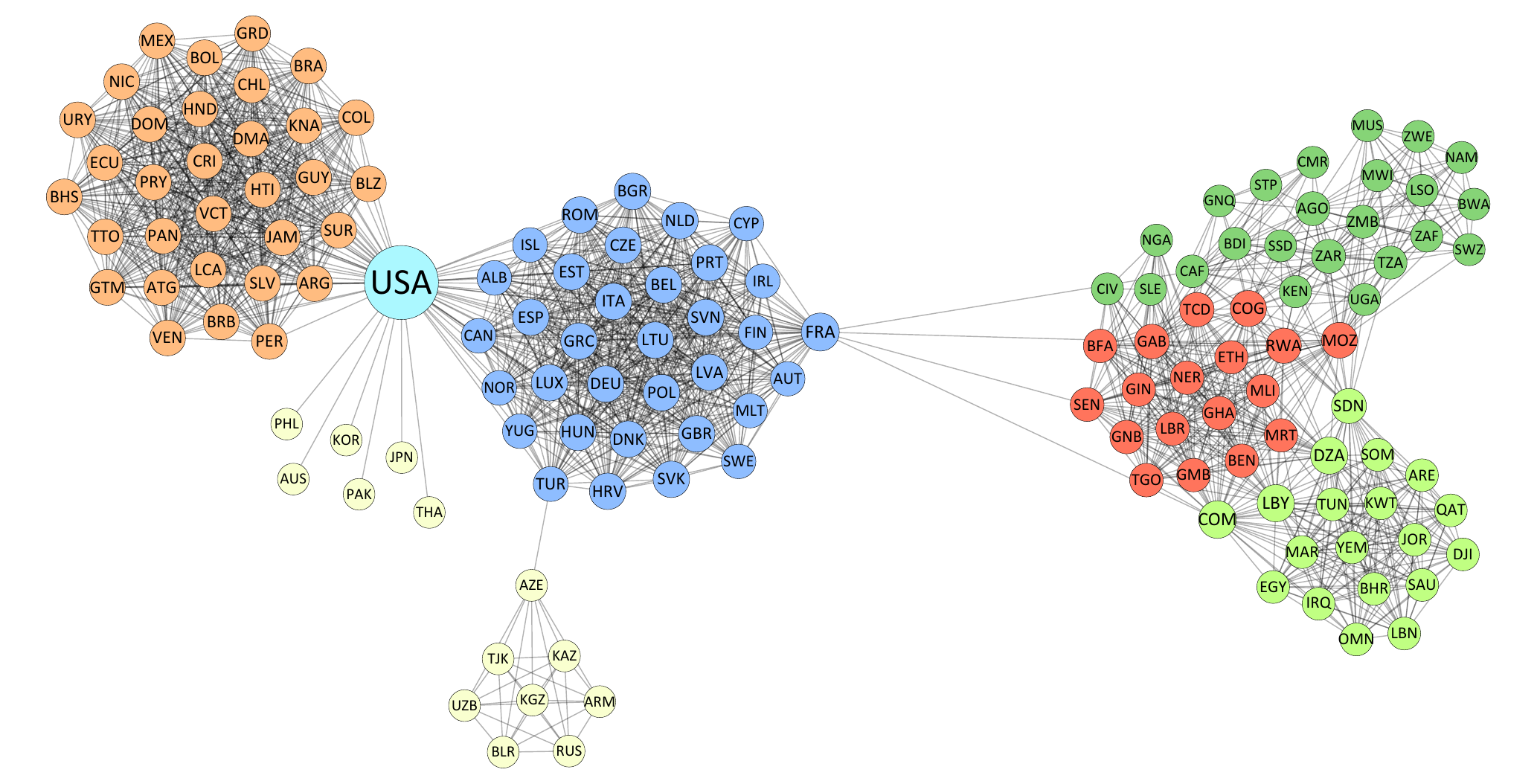}
	\caption{Global network of political alliances in 2016. Two countries are connected if they have taken part in a strong alliance treaty. Labels indicate country codes, while nodes are coloured by block memberships found through the standard stochastic blockmodel.}
	\label{fig1}
\end{figure}

\begin{figure}[tbh]
	\centering
	\includegraphics[width=\linewidth,trim=0cm 1cm 0cm 0cm]{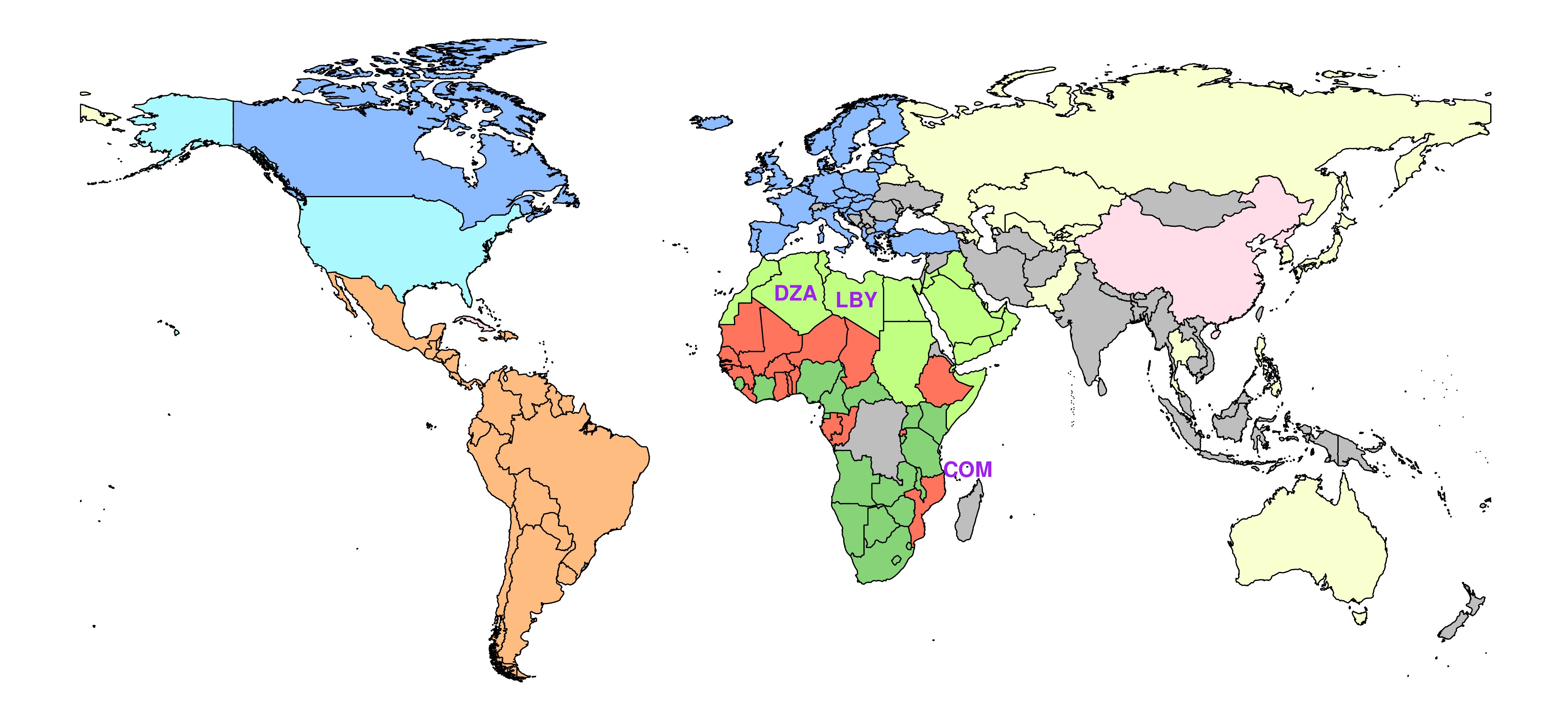}
	\caption{World map with countries coloured by block. Colours are kept consistent with Figure \ref{fig1}. Countries coloured in grey are isolates, meaning that they were not part of any strong military alliance as of 2016. The isolated group formed by China, Cuba and North Korea is coloured in pink. Purple labels indicate  the countries with the highest uncertainty in block-membership.}
	\label{figmap}
\end{figure}

To model the network, we use the standard version of the stochastic blockmodel, as in (\ref{eq:sbm}). In this case, we have opted for a relatively small number of communities (seven to be precise) to try and capture the larger blocks of military alliances. 
Estimation was performed using the Monte-Carlo-based EM routine. The resulting fitted block decomposition is given in Figure \ref{fig1}. The associated world map is shown in Figure \ref{figmap}, where countries are coloured by block. States coloured in grey on the map are isolates in the network, meaning that they are not involved in any strong military alliance in 2016. Moreover, China, Cuba and North Korea (coloured in pink) are only connected to each other, and are thus isolated from the rest of the network and therefore excluded from the model fitting.
The plots show how the blocks recovered by the stochastic blockmodel are very much related and in accordance with the geopolitical structure of the modern world. The network can be visually split into two large components.  In the first component, on the left side of the plot in Figure \ref{fig1}, the central blue block contains most European countries together with Canada. This block is very densely linked, as most of the countries inside it belong to NATO and other major alliances. The orange block pretty much coincides with Central and South-America, and it is also quite dense. The European and the American block are linked by the USA, which, given its unique connectivity behaviour, constitutes a block on its own. The yellow block includes mostly Asiatic countries as well as some Pacific states.
The other component of the network, on the right side of Figure~\ref{fig1}, is made out of three blocks. The light green block contains all countries from the Middle East together with Northern African countries such as Libya, Tunisia, Egypt and Morocco. The red block includes countries from Central and Western Africa. Finally, the dark green block is composed of Southern African countries. This Southern block and the Northern African one are bridged by Sudan. As an additional note, we can observe that the two major components of the network are linked exclusively through France, that, while belonging to the European block, acts as a bridge between Africa and Europe itself.

\begin{figure}[tbh]
	\centering
	\includegraphics[width=0.8\linewidth]{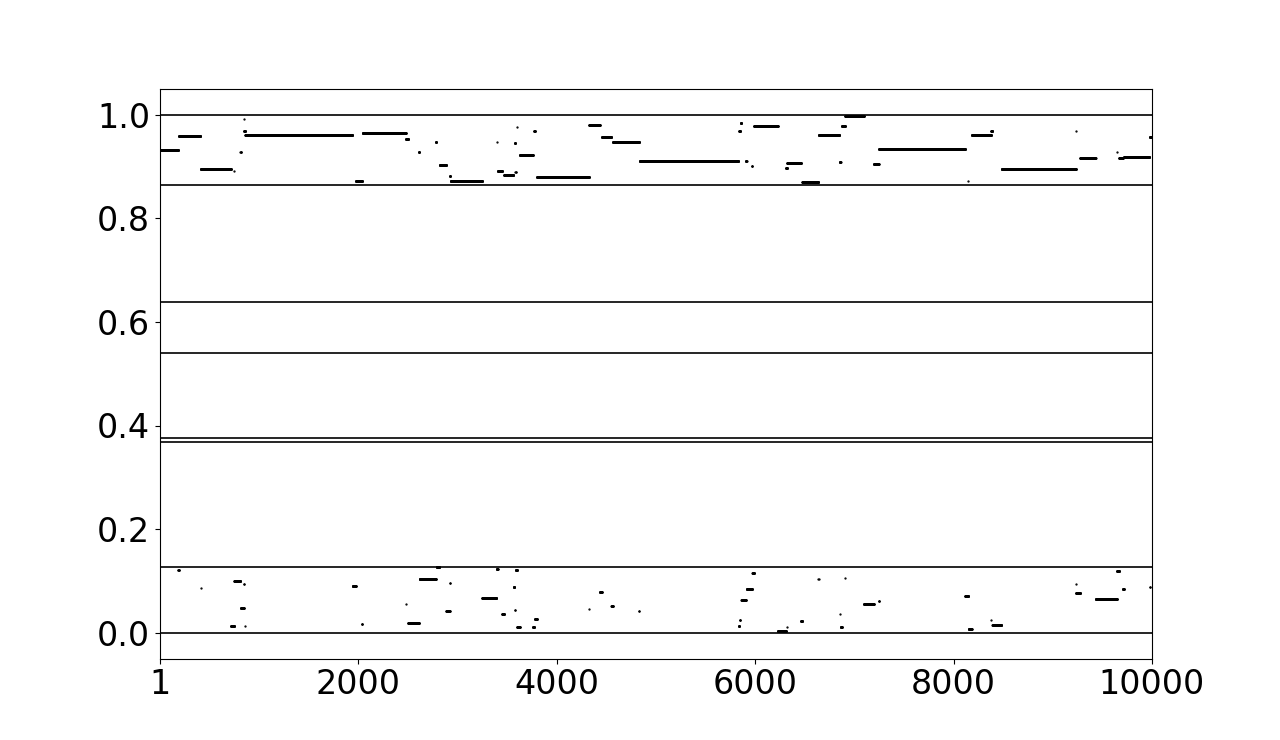}
	\caption{Posterior sample of the latent quantity $U$ for Libya plotted against the number of MCMC stages. Horizontal lines represent community boundaries.}
	\label{fig:postSamp}
\end{figure}

In addition to the block structure, we also investigate the uncertainty of the node allocation, using the Monte-Carlo-based posterior samples. We therefore consider the last Gibbs sampling sequence after the algorithm has converged. We investigate the three countries with the lowest values of the normalised Gini coefficient calculated over the allocation frequencies, which in turn implies the highest uncertainty. These countries are Libya (LBY), Algeria (DZA) and Comoros (COM), which all belong the light green Arabic block. The switching of communities exhibited by Lybia throughout the posterior sampling is illustrated as an example in Figure~\ref{fig:postSamp}. 
It shows that the sample for $u_{\mbox{\textit{Lybia}}}$ mostly appears within $[0.87,1]$ (the interval of the light green block), but also exhibits some stages where it is within $[0,0.13]$ (the interval of the red block). The posterior frequencies for Libya as well as for Comoros and Algeria with respect to the different groups are shown in Table~\ref{tab:postFreq}, which also comprises the corresponding Gini coefficients. 
\begin{table}
\resizebox{\textwidth}{!}{\begin{tabular}{c|ccccccc|c}
	\multirow{2}{*}{Country} & \multicolumn{7}{c|}{Community} & Gini \\[-0.1cm]
	& red & blue & cyan & dark green & yellow & orange &  light green & coefficient \\\hline\hline
	Comoros	& 0.1748 & 0 & 0 & 0 & 0 & 0 & 0.8252 & 0.9417 \\
	Libya & 0.1598 & 0 & 0 & 0 & 0 & 0 & 0.8402 & 0.9467 \\
	Algeria & 0.1558 & 0 & 0 & 0 & 0 & 0 &0.8442 & 0.9481 \\
\end{tabular}}
\caption{Posterior frequencies for the three countries with the highest uncertainty in their community memberships. The corresponding normalised Gini coefficient is depicted in the rightmost column.}
\label{tab:postFreq}
\end{table}
The table shows how all three countries have a substantial tendency to move to the Central and Western African block. According to the fitted blockmodel, in 15 to 18 percent of the MCMC sample stages the three countries are assigned to this red block. Turning our attention to all other countries, we observe Gini coefficients which are close to one and thus exhibit only very little uncertainty in block-membership. Altogether, this reveals how the estimated community structure appears to be quite strong. 


\subsection{Butterfly Similarity Network}

\begin{figure}[t]
	\FloatBarrier
	\centering
	\captionsetup[subfigure]{labelformat=empty}
	\begin{subfigure}[c]{0.47\textwidth}
		\center
		\includegraphics[width=6.5cm,height=7cm,trim=3.9cm 2cm 2cm 3cm]{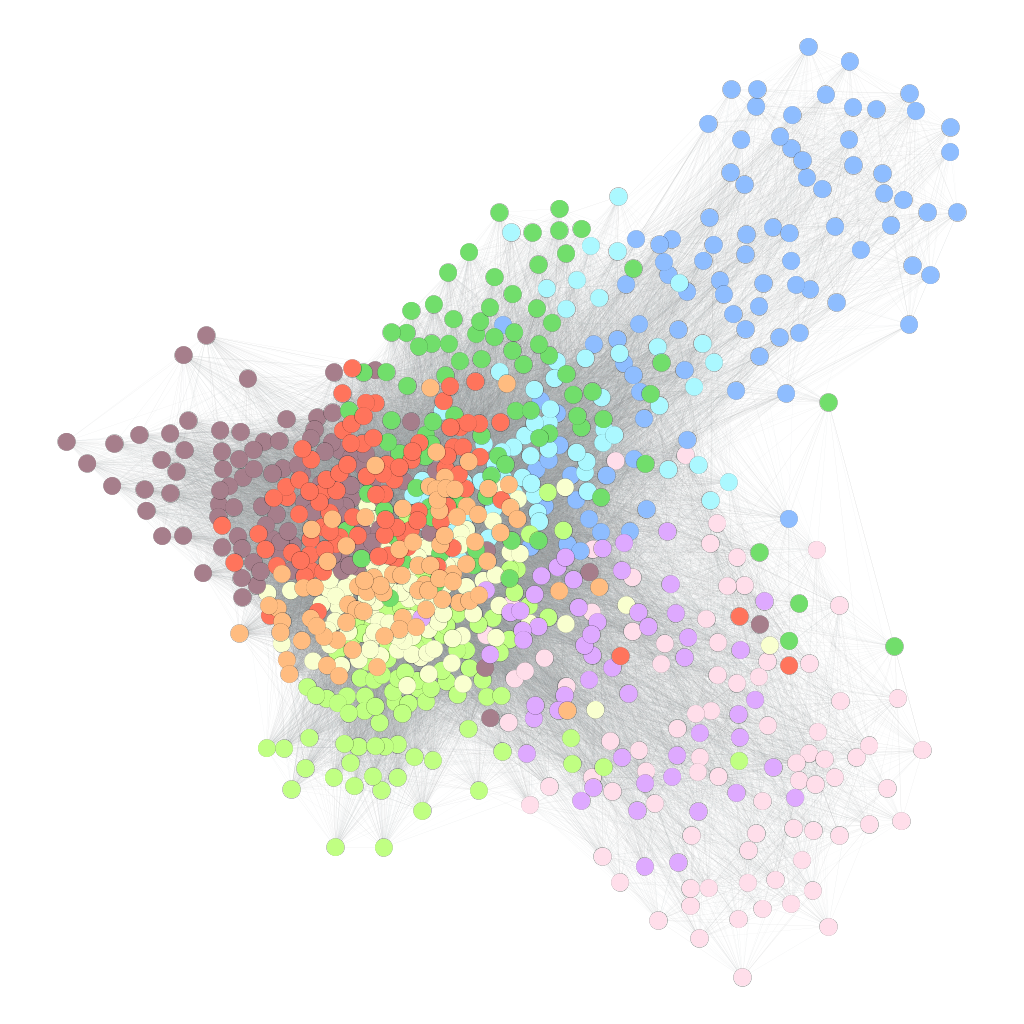}
		\subcaption{(a) Species}
		\FloatBarrier
	\end{subfigure}
	\begin{subfigure}[c]{0.47\textwidth}
		\center
		\includegraphics[width=6.5cm,height=7cm,trim=3cm 2cm 3.9cm 3cm]{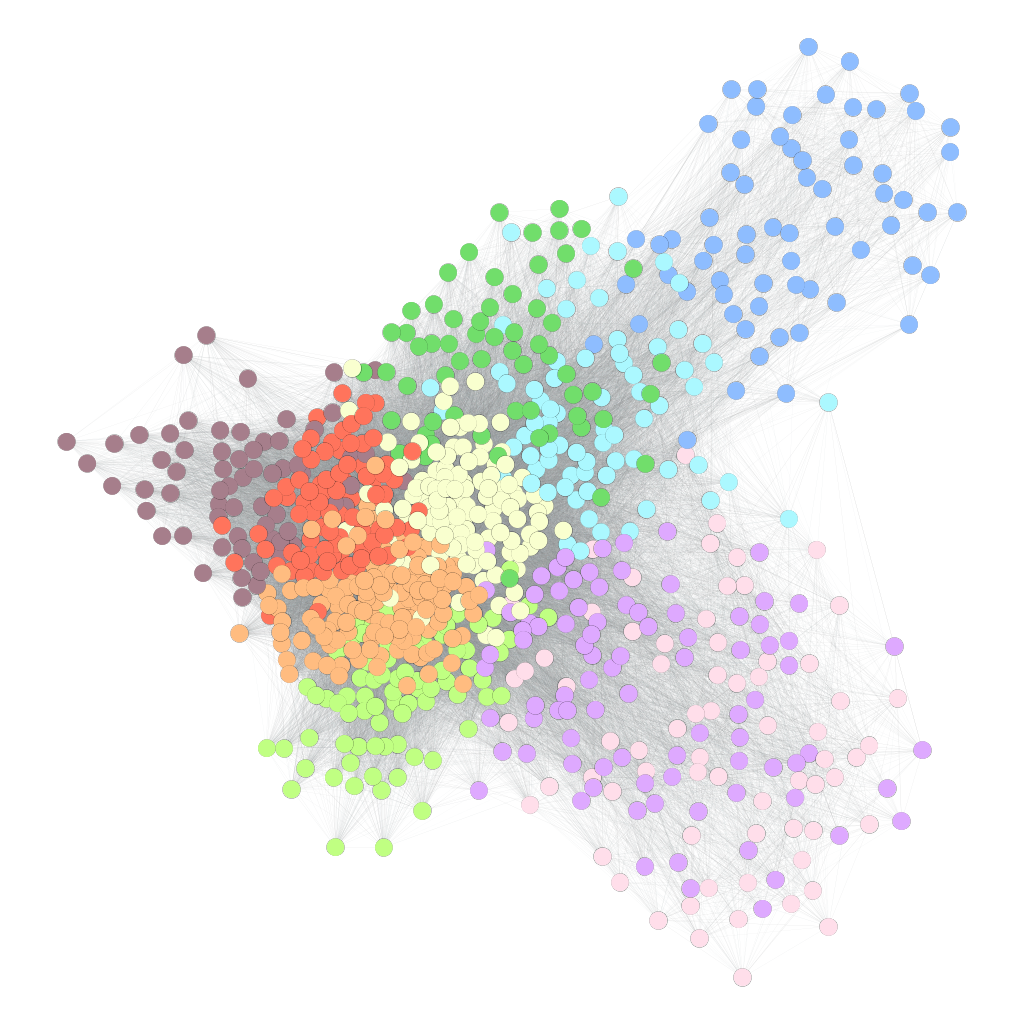}
		\subcaption{(b) Found Communities}
	\end{subfigure}
	\caption{Comparison between ``ground truth'' communities (species) and groups found by the poisson stochastic blockmodel in a network of butterflies, with weighted edges representing the degree of visual similarity between them.}
	\label{fig2}
\end{figure}

To model this network, we use the Poisson version of the standard stochastic blockmodel as in (\ref{eq:4}), taking advantage of the fact that this variant is suitable to treat multi-edged networks as well as binary ones. To fit this model, underlying similarity measures were discretized into count data through binning. Estimation was performed using the Variational EM approach developed by \cite{Mariadassou2010} and implemented in R by \cite{Leger2016}.
In this case, since we know that the real number of species is ten, we can simply use the same number of communities for the estimation. Figure \ref{fig2} shows the results of the model fit compared with the partition of butterflies into species.

At a first glance, we can see that the communities recovered mirror the real species relatively well. The most evident difference is the fact that two species (orange and yellow) are apparently really similar, and are therefore split up by the blockmodel. Other than that, the structure that was found does not appear to present major differences from the biological classification of the species. To quantify the goodness of the recovered block-structure compared to the ``ground-truth" communities, several measures are available (we refer to \cite{Jebabli2018} for a comprehensive survey). Here we opted for the Rand Index, a measure of similarity between two data clusterings that can simply be described as the number of agreements in classifying pairs divided by the total number of pairs \citep{Rand1971}. 
The index takes values between $0$ and $1$, and in this case it is equal to $0.91$, indicating that, given two Butterflies chosen at random, the blockmodel is able to correctly identify if they belong to the same species or not 91\% of times. 


\subsection{E-Mail Exchanges Network}

\begin{figure}[h!]
	\FloatBarrier
	\centering
	\captionsetup[subfigure]{labelformat=empty}
	\begin{subfigure}[c]{0.47\textwidth}
		\center
		\includegraphics[width=6.8cm,trim=0.7cm 2cm 0cm 1.5cm]{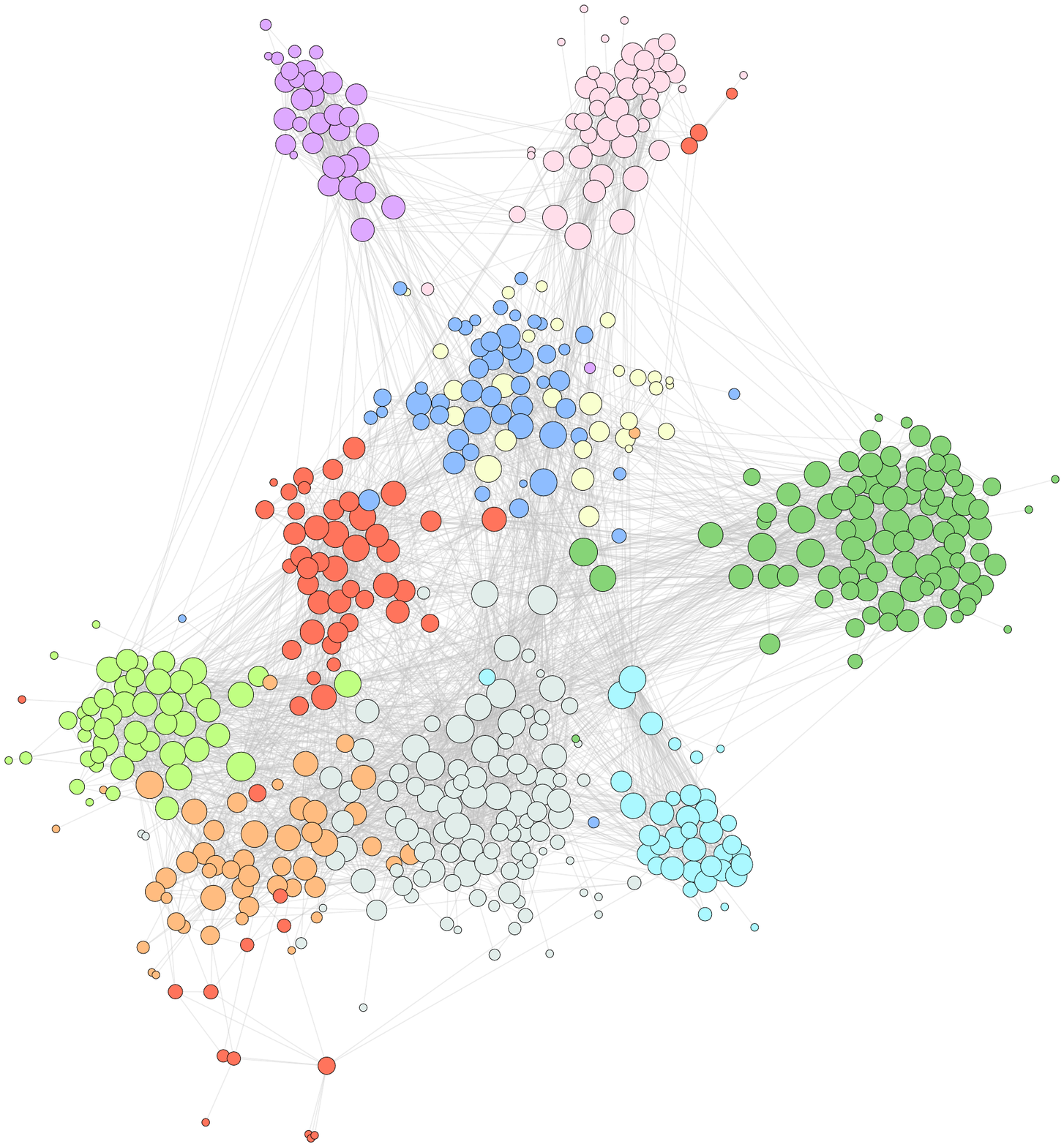}
		\subcaption{(a) Real Departments}
		\FloatBarrier
	\end{subfigure}
	\begin{subfigure}[c]{0.47\textwidth}
		\center
		\includegraphics[width=6.8cm,trim=0cm 2cm 1cm 1.5cm]{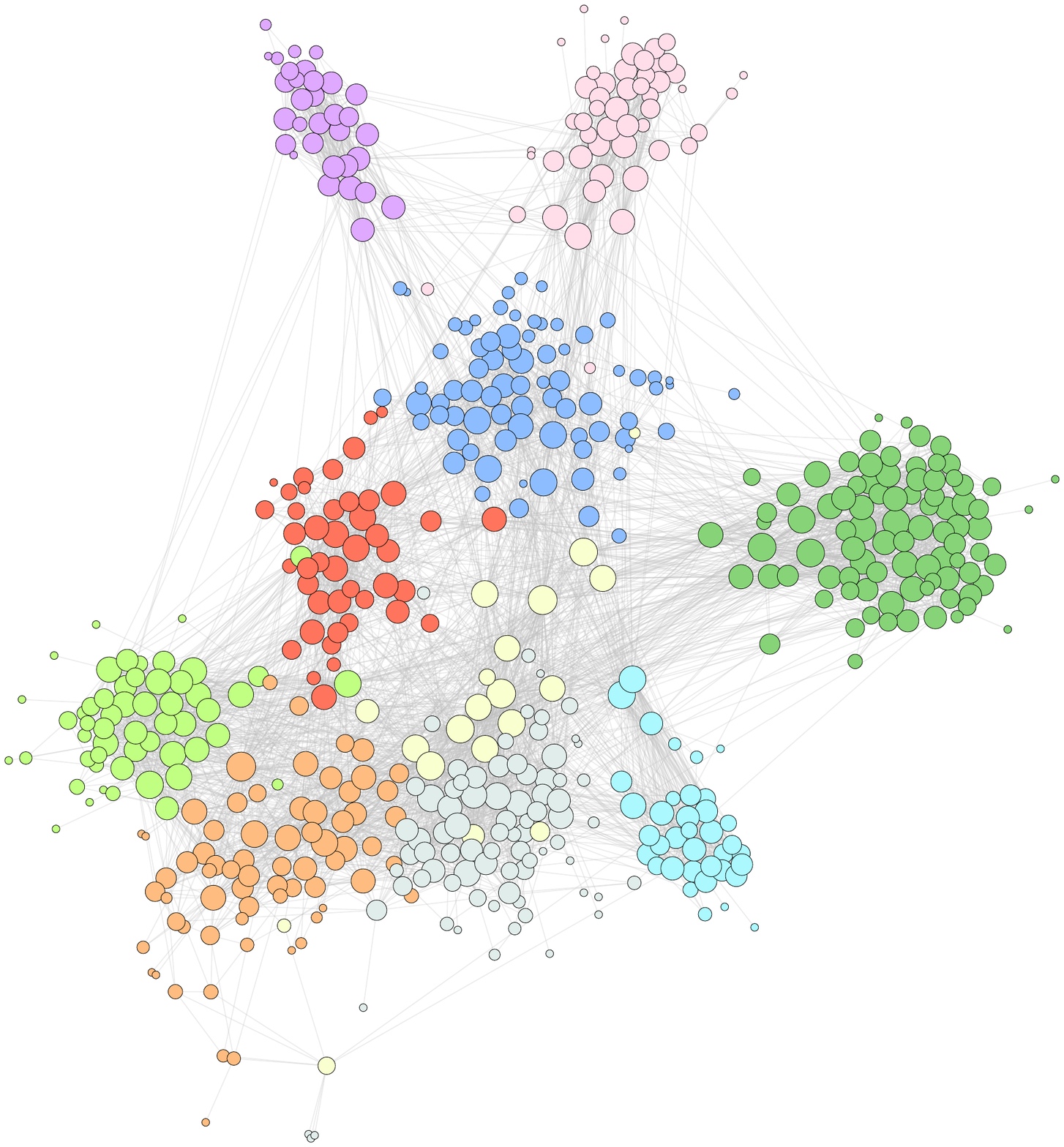}
		\subcaption{(b) Found Communities}
	\end{subfigure}
	\caption{Comparison between ``ground truth'' communities (departments) and groups found by the degree-corrected stochastic blockmodel in a network of e-mail exchanges within a large European research institution.}
	\label{fig3}
\end{figure}

The network of e-mails within a research institution exhibits a skewed degree distribution that is typical of social networks. 
As explained above, one way to circumvent this issue is to use degree-correction. For this application, we therefore made use of the original version of the degree-corrected stochastic blockmodel as in (\ref{eq:degcor}) \citep{Karrer2011}.
%
The results of the model fitting, together with the partitioning of the network into real departments, are visualized in Figure \ref{fig3}.
Looking at the plots, it is evident how the model with degree-correction is able to recover the communities quite accurately. Comparing the partition discovered by the SBM with the actual departments, one small department (yellow) merges into another one (blue), and an additional community is therefore found in the central area of the network, splitting a department (grey) into two. Other than that, the structure found is remarkably similar to the partition induced by the departments, with some exceptions due to the existence of disconnected components within departments. In this case the Rand Index is equal to 0.95, indicating a very high level of agreement among the partitions.


\section{Conclusions}
\label{sec:disc}

Mixture modelling can be extended to network data through stochastic blockmodels. Networks are rather complex structures, leading to computationally demanding estimation routines. Several algorithms specific for this class of problems have emerged over time, some of which are discussed in this paper. We also provided an overview of different types of blockmodels by applying them to real-world network datasets.
Among others, one of the models that we showcased is the degree-corrected stochastic blockmodel, which is particularly well-suited for networks with a highly skewed degree structure. 

Considering stochastic blockmodels (and community detection problems) as mixture models opens up a new avenue of extensions and novel models. Looking at the many model proposals in the field of mixture, ranging from mixing different distributions towards the mixture of experts, it is evident that these extensions can be brought forward in network modelling with mixtures as well.  In fact, block-wise constant connectivity probabilities could be extended towards non-constant ones. Moreover, covariates could also be included. These extensions lie well beyond the scope of this paper, but it is evident how the long history of mixture models, which started with \cite{Pearson:94}, has not come to an end, and extends promisingly in the realm of networks.


\section*{Acknowledgements}
We would like to thank the European Cooperation in Science and Technology [COST Action CA15109 (COSTNET)]. The first author would also like to thank Cornelius Fritz for many meaningful comments and discussions. Finally, the last author would like to thank Murray Aitkin for his enthusiasm, ingenuity and open-mindedness with respect to statistics, and, last but not least, for his friendship.

\section*{Appendix}

\subsection*{Details on MCMC Sampling}


Assuming $\boldsymbol{u}^{<t>} = (u_1^{<t>}, \ldots, u_N^{<t>}) $ to be the current state of the Markov chain, we 
update the $j$-th component as follows. At first, we set $u_l^{<t+1>} = u_l^{<t>}$ for $l \neq j$, while for component $u_j$ we draw a new potential state $u_j^*$ from a uniform proposal with regard to $[0,1] \setminus [\tau_{k(j,<t>)-1}, \tau_{k(j,<t>)})$ and with $[\tau_{k(j,<t>)-1}, \tau_{k(j,<t>)})$ being the sub-interval that includes $u_j^{<t>}$. This leads to the acceptance probability
\begin{multline}
\min \left\{ 1, \; \prod_{l\neq j}^N \left[ \left( \frac{p(u_j^*, u_l^{<t>})}{p (u_j^{<t>}, u_l^{<t>})} \right)^{y_{ij}} \left( \frac{1- p(u_j^*, u_l^{<t>})}{1- p (u_j^{<t>}, u_l^{<t>})} \right)^{1-y_{ij}} \right] \right. \\
\left. \vphantom{\prod_{l\neq j}^N \left[ \left( \frac{p(u_j^*, u_l^{<t>})}{p (u_j^{<t>}, u_l^{<t>})} \right)^{y_{ij}} \left( \frac{1- p(u_j^*, u_l^{<t>})}{1- p (u_j^{<t>}, u_l^{<t>})} \right)^{1-y_{ij}} \right]} \cdot \frac{1 - (\tau_{k(j,<t>)} - \tau_{k(j,<t>)-1})}{1 - (\tau_{k(j,*)} - \tau_{k(j,*)-1})} \right\},
\label{eq:accProb}
\end{multline}
where $[\tau_{k(j,*)-1}, \tau_{k(j,*)})$ represents the sub-interval which includes $u_j^*$. In the event of rejection we remain with the previous value $u_j^{<t>}$. Running the Markov chain, we get a simulation-based estimate of the group mode, which concludes the E step. It should be mentioned that in the beginning the number of Gibbs sampling stages taken into account for approximating the mode can be rather small, since the early model configurations are potentially far from the truth and thus already imply a deviating reallocation. 



\bibliographystyle{chicago}

\bibliography{ms}

\begin{thebibliography}{}

\bibitem[\protect\citeauthoryear{Abbe}{Abbe}{2018}]{Abbe:19}
Abbe, E. (2018).
\newblock {Community detection and stochastic block models}.
\newblock {\em Foundations and Trends in Communications and Information
  Theory\/}~{\em 14\/}(1-2), 1--162.

\bibitem[\protect\citeauthoryear{Airoldi, Blei, Fienberg, and Xing}{Airoldi
  et~al.}{2008}]{Airoldi2008}
Airoldi, E.~M., D.~M. Blei, S.~E. Fienberg, and E.~P. Xing (2008).
\newblock {Mixed Membership Stochastic Blockmodels.}
\newblock {\em Journal of machine learning research\/}~{\em 9}, 1981--2014.

\bibitem[\protect\citeauthoryear{Airoldi, Costa, and Chan}{Airoldi
  et~al.}{2013}]{Airoldi:13}
Airoldi, E.~M., T.~B. Costa, and S.~H. Chan (2013).
\newblock {Stochastic blockmodel approximation of a graphon: Theory and
  consistent estimation}.
\newblock {\em Advances in Neural Information Processing Systems\/}, 692--700.

\bibitem[\protect\citeauthoryear{Aitkin}{Aitkin}{1980}]{Aitkin:80}
Aitkin, M. (1980).
\newblock {Mixture applications of the EM algorithm in GLIM}.
\newblock {\em COMPSTAT\/}, 537--541.

\bibitem[\protect\citeauthoryear{Aitkin}{Aitkin}{2011}]{Aitkin2011}
Aitkin, M. (2011).
\newblock {\em {How many components in a finite mixture?}}
\newblock Wiley.

\bibitem[\protect\citeauthoryear{Aitkin and Rubin}{Aitkin and
  Rubin}{1985}]{AitkinRubin:85}
Aitkin, M. and D.~B. Rubin (1985).
\newblock {Estimation and Hypothesis Testing in Finite Mixture Models}.
\newblock {\em Journal of the Royal Statistical Society: Series B\/}~{\em
  47\/}(1), 67--75.

\bibitem[\protect\citeauthoryear{Aitkin, Vu, and Francis}{Aitkin
  et~al.}{2014}]{Aitkin:14}
Aitkin, M., D.~Vu, and B.~Francis (2014).
\newblock {Statistical modelling of the group structure of social networks}.
\newblock {\em Social Networks\/}~{\em 38}, 74--87.

\bibitem[\protect\citeauthoryear{Aitkin and Wilson}{Aitkin and
  Wilson}{1980}]{AitkinWilson:80}
Aitkin, M. and T.~G. Wilson (1980).
\newblock {Mixture Models, outliers and the EM algorithm}.
\newblock {\em Technometrics\/}~{\em 22}, 325--331.

\bibitem[\protect\citeauthoryear{Barab{\'{a}}si and Albert}{Barab{\'{a}}si and
  Albert}{1999}]{Barabasi1999}
Barab{\'{a}}si, A.~L. and R.~Albert (1999).
\newblock {Emergence of scaling in random networks}.
\newblock {\em Science\/}~{\em 286\/}(5439), 509--512.

\bibitem[\protect\citeauthoryear{Benaglia, Chauveau, Hunter, and
  Young}{Benaglia et~al.}{2009}]{Benagalia:09}
Benaglia, T., D.~Chauveau, D.~R. Hunter, and D.~S. Young (2009).
\newblock {mixtools: An R Package for Analyzing Mixture Models}.
\newblock {\em Journal of Statistical Software\/}~{\em 32\/}(6).

\bibitem[\protect\citeauthoryear{Biagini, Kauermann, and Meyer-Brandis}{Biagini
  et~al.}{2019}]{BiageniKauermannMeye:19}
Biagini, F., G.~Kauermann, and T.~Meyer-Brandis (2019).
\newblock {\em {Network Science}}.
\newblock Springer-Verlag.

\bibitem[\protect\citeauthoryear{B{\"{o}}hning}{B{\"{o}}hning}{1999}]{Boehning:99}
B{\"{o}}hning, D. (1999).
\newblock {\em {Computer Assisted Analysis of Mixtures and Applications: Meta
  Analysis, Disease Mapping and Others}}.
\newblock Chapman and Hall/CRC.

\bibitem[\protect\citeauthoryear{Bui, Chaudhuri, Leighton, and Sipser}{Bui
  et~al.}{1987}]{Bui:87}
Bui, T.~N., S.~Chaudhuri, F.~T. Leighton, and M.~Sipser (1987).
\newblock {Graph bisection algorithms with good average case behavior}.
\newblock {\em Combinatorica\/}~{\em 7}, 171--191.

\bibitem[\protect\citeauthoryear{Chen and Lei}{Chen and Lei}{2018}]{Chen2018}
Chen, K. and J.~Lei (2018).
\newblock {Network Cross-Validation for Determining the Number of Communities
  in Network Data}.
\newblock {\em Journal of the American Statistical Association\/}~{\em
  113\/}(521), 241--251.

\bibitem[\protect\citeauthoryear{Dempster, Laird, and Rubin}{Dempster
  et~al.}{1977}]{DempLairdRubin:77}
Dempster, A., N.~Laird, and D.~Rubin (1977).
\newblock {Maximum likelihood from incomplete observations}.
\newblock {\em Journal of the Royal Statistical Society, Series B\/}~{\em 39},
  1--38.

\bibitem[\protect\citeauthoryear{Everitt and Hand}{Everitt and
  Hand}{1980}]{EverittHand:80}
Everitt, B.~S. and D.~J. Hand (1980).
\newblock {\em {Finite mixture distributions}}.
\newblock Chapman {\&} Hall.

\bibitem[\protect\citeauthoryear{Fienberg}{Fienberg}{2012}]{Fienberg:12}
Fienberg, S.~E. (2012).
\newblock {A brief history of statistical models for network analysis and open
  challenges}.
\newblock {\em Journal of Computational and Graphical Statistics\/}~{\em
  21\/}(4), 825--839.

\bibitem[\protect\citeauthoryear{Friedl and Kauermann}{Friedl and
  Kauermann}{2000}]{Friedl2000}
Friedl, H. and G.~Kauermann (2000).
\newblock {Standard errors for EM estimates in generalized linear models with
  random effects}.
\newblock {\em Biometrics\/}~{\em 56\/}(3), 761--767.

\bibitem[\protect\citeauthoryear{Fr{\"{u}}hwirth-Schnatter}{Fr{\"{u}}hwirth-Schnatter}{2006}]{Fruehwirth:06}
Fr{\"{u}}hwirth-Schnatter, S. (2006).
\newblock {\em {Finite Mixture and Markov Switching Models}}.
\newblock Springer-Verlag.

\bibitem[\protect\citeauthoryear{Fr{\"{u}}hwirth-Schnatter, Celeux, and
  Robert}{Fr{\"{u}}hwirth-Schnatter et~al.}{2019}]{Fruehwirth:19}
Fr{\"{u}}hwirth-Schnatter, S., G.~Celeux, and C.~P. Robert (2019).
\newblock {\em {Handbook of Mixture Analysis}}.
\newblock Chapman {\&} Hall.

\bibitem[\protect\citeauthoryear{Goldenberg, Zheng, Fienberg, and
  Airoldi}{Goldenberg et~al.}{2009}]{Goldenberg:10}
Goldenberg, A., A.~X. Zheng, S.~E. Fienberg, and E.~M. Airoldi (2009).
\newblock {A survey of statistical network models}.
\newblock {\em Foundations and Trends in Machine Learning\/}~{\em 2\/}(2),
  129--233.

\bibitem[\protect\citeauthoryear{Gormley and Fr{\"{u}}hwirth-Schnatter}{Gormley
  and Fr{\"{u}}hwirth-Schnatter}{2019}]{GormFrueh:19}
Gormley, I.~C. and S.~Fr{\"{u}}hwirth-Schnatter (2019).
\newblock {Mixture of Experts Models}.
\newblock In {\em Handbook of Mixture Analysis}, pp.\  271--307. Chapman and
  Hall/CRC.

\bibitem[\protect\citeauthoryear{Gormley and Murphy}{Gormley and
  Murphy}{2010}]{GormleyMurphy:10}
Gormley, I.~C. and T.~B. Murphy (2010).
\newblock {A mixture of experts latent position cluster model for social
  network data}.
\newblock {\em Statistical Methodology\/}~{\em 7\/}(3), 385--405.

\bibitem[\protect\citeauthoryear{Gupta and Huang}{Gupta and
  Huang}{1981}]{GuptaHuany:81}
Gupta, S.~S. and W.~T. Huang (1981).
\newblock {On mixture of distributions: A survey and some new results on
  ranking and selection}.
\newblock {\em Sankhya: The indian journal of Statistics\/}~{\em 43}, 245--290.

\bibitem[\protect\citeauthoryear{Holland, Laskey, and Leinhardt}{Holland
  et~al.}{1983}]{HollandLaskeyLein:83}
Holland, P.~W., K.~Laskey, and S.~Leinhardt (1983).
\newblock {Stochastic blockmodels: First steps}.
\newblock {\em Social Networks\/}~{\em 5}, 109--137.

\bibitem[\protect\citeauthoryear{Hunter, Handcock, Butts, Goodreau, and
  Morris}{Hunter et~al.}{2008}]{Hunter:12}
Hunter, D.~R., M.~S. Handcock, C.~T. Butts, S.~M. Goodreau, and M.~Morris
  (2008).
\newblock {ergm: A package to fit, simulate and diagnose exponential-family
  models for networks}.
\newblock {\em Journal of Statistical Software\/}~{\em 24\/}(3).

\bibitem[\protect\citeauthoryear{Jacobs, Jordan, Nowlan, and Hinton}{Jacobs
  et~al.}{1991}]{Jacobs:91}
Jacobs, R.~A., M.~I. Jordan, S.~J. Nowlan, and G.~E. Hinton (1991).
\newblock {Adaptive mixtures of local experts}.
\newblock {\em Neural computation\/}~{\em 3\/}(1), 79--87.

\bibitem[\protect\citeauthoryear{Jebabli, Cherifi, Cherifi, and
  Hamouda}{Jebabli et~al.}{2018}]{Jebabli2018}
Jebabli, M., H.~Cherifi, C.~Cherifi, and A.~Hamouda (2018).
\newblock {Community detection algorithm evaluation with ground-truth data}.
\newblock {\em Physica A: Statistical Mechanics and its Applications\/}~{\em
  492}, 651--706.

\bibitem[\protect\citeauthoryear{Jordan, Ghahramani, Jaakkola, and Saul}{Jordan
  et~al.}{1999}]{Jordan1999}
Jordan, M.~I., Z.~Ghahramani, T.~S. Jaakkola, and L.~K. Saul (1999).
\newblock {Introduction to variational methods for graphical models}.
\newblock {\em Machine Learning\/}~{\em 37\/}(2), 183--233.

\bibitem[\protect\citeauthoryear{Karrer and Newman}{Karrer and
  Newman}{2011}]{Karrer2011}
Karrer, B. and M.~E. Newman (2011).
\newblock {Stochastic blockmodels and community structure in networks}.
\newblock {\em Physical Review E - Statistical, Nonlinear, and Soft Matter
  Physics\/}~{\em 83\/}(1), 016107.

\bibitem[\protect\citeauthoryear{Kauermann and Sischka}{Kauermann and
  Sischka}{2019}]{kauermann2019}
Kauermann, G. and B.~Sischka (2019).
\newblock {Bayesian and Spline based Approaches for (EM based) Graphon
  Estimation}.
\newblock {\em arXiv:1903.06936\/}.

\bibitem[\protect\citeauthoryear{Kernighan and Lin}{Kernighan and
  Lin}{1970}]{Kernighan1970}
Kernighan, B.~W. and S.~Lin (1970).
\newblock {An Efficient Heuristic Procedure for Partitioning Graphs}.
\newblock {\em Bell System Technical Journal\/}~{\em 49\/}(2), 291--307.

\bibitem[\protect\citeauthoryear{Kolaczyk}{Kolaczyk}{2009}]{Kolaczyk:09}
Kolaczyk, E.~D. (2009).
\newblock {\em {Statistical Analysis of Network Data: Methods and Models}}.
\newblock Springer.

\bibitem[\protect\citeauthoryear{Kolaczyk and Cs{\'{a}}rdi}{Kolaczyk and
  Cs{\'{a}}rdi}{2014}]{KolaczykCsardi:14}
Kolaczyk, E.~D. and G.~Cs{\'{a}}rdi (2014).
\newblock {\em {Statistical Analysis of Network Data with R}}.
\newblock Springer.

\bibitem[\protect\citeauthoryear{Latouche and Robin}{Latouche and
  Robin}{2016}]{latouche2016}
Latouche, P. and S.~Robin (2016).
\newblock {Variational Bayes model averaging for graphon functions and motif
  frequencies inference in W-graph models}.
\newblock {\em Statistics and Computing\/}~{\em 26\/}(6), 1173--1185.

\bibitem[\protect\citeauthoryear{Lee, Xue, and Hunter}{Lee
  et~al.}{2020}]{Lee2020}
Lee, K.~H., L.~Xue, and D.~R. Hunter (2020).
\newblock {Model-based clustering of time-evolving networks through temporal
  exponential-family random graph models}.
\newblock {\em Journal of Multivariate Analysis\/}~{\em 175}, 104540.

\bibitem[\protect\citeauthoryear{Leeds, Ritter, Mitchell, and Long}{Leeds
  et~al.}{2002}]{Leeds2002}
Leeds, B.~A., J.~M. Ritter, S.~M.~L. Mitchell, and A.~G. Long (2002).
\newblock {Alliance treaty obligations and provisions, 1815-1944}.
\newblock {\em International Interactions\/}~{\em 28\/}(3), 237--260.

\bibitem[\protect\citeauthoryear{Leger}{Leger}{2016}]{Leger2016}
Leger, J.-B. (2016).
\newblock {Blockmodels: A R-package for estimating in Latent Block Model and
  Stochastic Block Model, with various probability functions, with or without
  covariates}.
\newblock {\em arXiv:1602.07587\/}.

\bibitem[\protect\citeauthoryear{Leisch}{Leisch}{2004}]{Leisch:04}
Leisch, F. (2004).
\newblock {FlexMix: A general framework for finite mixture models and latent
  class regression in R}.
\newblock {\em Journal of Statistical Software\/}~{\em 11}, 1--18.

\bibitem[\protect\citeauthoryear{Leskovec and Krevl}{Leskovec and
  Krevl}{2014}]{snapnets}
Leskovec, J. and A.~Krevl (2014).
\newblock {SNAP Datasets: Stanford Large Network Dataset Collection}.

\bibitem[\protect\citeauthoryear{Lindsay}{Lindsay}{1995}]{Lindsay:95}
Lindsay, B.~G. (1995).
\newblock {\em {Mixture Models: Theory, Geometry and Applications}}.
\newblock NSF-CVMS Regional Conference Series in Probability and Statistics. 5.

\bibitem[\protect\citeauthoryear{Lusher, Koskinen, and Robins}{Lusher
  et~al.}{2013}]{Lusker:13}
Lusher, D., J.~Koskinen, and G.~Robins (2013).
\newblock {Exponential random graph models for social networks: Theory,
  methods, and applications}.
\newblock {\em Cambridge University Press\/}~{\em 123\/}(1).

\bibitem[\protect\citeauthoryear{Mariadassou, Robin, and Vacher}{Mariadassou
  et~al.}{2010}]{Mariadassou2010}
Mariadassou, M., S.~Robin, and C.~Vacher (2010).
\newblock {Uncovering latent structure in valued graphs: A variational
  approach}.
\newblock {\em Annals of Applied Statistics\/}~{\em 4\/}(2), 715--742.

\bibitem[\protect\citeauthoryear{Masoudnia and Ebrahimpour}{Masoudnia and
  Ebrahimpour}{2014}]{MasoudniaEbra:14}
Masoudnia, S. and R.~Ebrahimpour (2014).
\newblock {Mixture of experts: a literature survey}.
\newblock {\em Artificial Intelligence Review\/}~{\em 42\/}(2), 275--293.

\bibitem[\protect\citeauthoryear{McLachlan and Peel}{McLachlan and
  Peel}{2000}]{MacLachlanPeel:00}
McLachlan, G.~J. and D.~Peel (2000).
\newblock {\em {Finite Mixture Models}}.
\newblock Wiley.

\bibitem[\protect\citeauthoryear{Nowicki and Snijders}{Nowicki and
  Snijders}{2001}]{Nowicki2001}
Nowicki, K. and T.~A. Snijders (2001).
\newblock {Estimation and prediction for stochastic blockstructures}.
\newblock {\em Journal of the American Statistical Association\/}~{\em
  96\/}(455), 1077--1087.

\bibitem[\protect\citeauthoryear{Olhede and Wolfe}{Olhede and
  Wolfe}{2014}]{WolfeOl:14}
Olhede, S.~C. and P.~J. Wolfe (2014).
\newblock {Network histograms and universality of blockmodel approximation}.
\newblock {\em Proceedings of the National Academy of Sciences\/}~{\em
  111\/}(41), 14722--14727.

\bibitem[\protect\citeauthoryear{Pearson}{Pearson}{1894}]{Pearson:94}
Pearson, K. (1894).
\newblock {Contributions to the mathematical theory of evolution}.
\newblock {\em Philosophical Trans. Roy. Soc. London\/}~{\em 185}, 71--110.

\bibitem[\protect\citeauthoryear{Peixoto}{Peixoto}{2017}]{Peixoto2017}
Peixoto, T.~P. (2017).
\newblock {Nonparametric Bayesian inference of the microcanonical stochastic
  block model}.
\newblock {\em Physical Review E\/}~{\em 95\/}(1), 012317.

\bibitem[\protect\citeauthoryear{Rand}{Rand}{1971}]{Rand1971}
Rand, W.~M. (1971).
\newblock {Objective criteria for the evaluation of clustering methods}.
\newblock {\em Journal of the American Statistical Association\/}~{\em
  66\/}(336), 846--850.

\bibitem[\protect\citeauthoryear{Renshaw, Titterington, Smith, and
  Makov}{Renshaw et~al.}{1987}]{TitteringtonSmithMavov:85}
Renshaw, A.~E., D.~M. Titterington, A.~F.~M. Smith, and H.~E. Makov (1987).
\newblock {\em {Statistical Analysis of Finite Mixture Distributions.}}, Volume
  150.
\newblock Wiley.

\bibitem[\protect\citeauthoryear{Riolo, Cantwell, Reinert, and Newman}{Riolo
  et~al.}{2017}]{Riolo2017}
Riolo, M.~A., G.~T. Cantwell, G.~Reinert, and M.~E. Newman (2017).
\newblock {Efficient method for estimating the number of communities in a
  network}.
\newblock {\em Physical Review E\/}~{\em 96\/}(3), 032310.

\bibitem[\protect\citeauthoryear{Robbins}{Robbins}{1948}]{Robbins:48}
Robbins, H. (1948).
\newblock {Mixture of distirbutions}.
\newblock {\em The Annals of Mathematical Statistics\/}~{\em 19}, 360--369.

\bibitem[\protect\citeauthoryear{Simon}{Simon}{1955}]{Simon1955}
Simon, H.~A. (1955).
\newblock {On a Class of Skew Distribution Functions}.
\newblock {\em Biometrika\/}~{\em 42\/}(3-4), 425--440.

\bibitem[\protect\citeauthoryear{Snijders and Nowicki}{Snijders and
  Nowicki}{1997}]{Snijders1997}
Snijders, T.~A. and K.~Nowicki (1997).
\newblock {Estimation and prediction for stochastic blockmodels for graphs with
  latent block structure}.
\newblock {\em Journal of Classification\/}~{\em 14\/}(1), 75--100.

\bibitem[\protect\citeauthoryear{Tallberg}{Tallberg}{2005}]{Tallberg2005}
Tallberg, C. (2005).
\newblock {A Bayesian approach to modeling stochastic blockstructures with
  covariates}.
\newblock {\em Journal of Mathematical Sociology\/}~{\em 29\/}(1), 1--23.

\bibitem[\protect\citeauthoryear{Teicher}{Teicher}{1960}]{Teicher:60}
Teicher, H. (1960).
\newblock {On the mixture of distributions}.
\newblock {\em The Annals of Mathematical Statistics\/}~{\em 31}, 55--73.

\bibitem[\protect\citeauthoryear{Vu and Aitkin}{Vu and Aitkin}{2015}]{Vu2015}
Vu, D. and M.~Aitkin (2015).
\newblock {Variational algorithms for biclustering models}.
\newblock {\em Computational Statistics and Data Analysis\/}~{\em 89}, 12--24.

\bibitem[\protect\citeauthoryear{Vu, Hunter, and Schweinberger}{Vu
  et~al.}{2013}]{vu2013}
Vu, D.~Q., D.~R. Hunter, and M.~Schweinberger (2013).
\newblock {Model-based clustering of large networks}.
\newblock {\em The Annals of Applied Statistics\/}~{\em 7\/}(2), 1010.

\bibitem[\protect\citeauthoryear{Wang, Markert, and Everingham}{Wang
  et~al.}{2009}]{Wang2009}
Wang, J., K.~Markert, and M.~Everingham (2009).
\newblock {Learning models for object recognition from natural language
  descriptions}.
\newblock In {\em British Machine Vision Conference, BMVC 2009 - Proceedings}.
  British Machine Vision Association, BMVA.

\bibitem[\protect\citeauthoryear{Wang and Wong}{Wang and Wong}{1987}]{Wang1987}
Wang, Y.~J. and G.~Y. Wong (1987).
\newblock {Stochastic blockmodels for directed graphs}.
\newblock {\em Journal of the American Statistical Association\/}~{\em
  82\/}(397), 8--19.

\bibitem[\protect\citeauthoryear{Wang and Bickel}{Wang and
  Bickel}{2017}]{Wang2017}
Wang, Y.~X. and P.~J. Bickel (2017).
\newblock {Likelihood-based model selection for stochastic block models}.
\newblock {\em Annals of Statistics\/}~{\em 45\/}(2), 500--528.

\bibitem[\protect\citeauthoryear{White, Boorman, and Breiger}{White
  et~al.}{1976}]{White}
White, H.~C., S.~A. Boorman, and R.~L. Breiger (1976).
\newblock {Social structure from multiple networks. I. Blockmodels of roles and
  positions}.
\newblock {\em American Journal of Sociology\/}~{\em 81\/}(4), 730--780.

\bibitem[\protect\citeauthoryear{Zittnik, Sosi{\v{c}}, Maheshwari, and
  Leskovec}{Zittnik et~al.}{2018}]{biosnapnets}
Zittnik, M., R.~Sosi{\v{c}}, S.~Maheshwari, and J.~Leskovec (2018).
\newblock {BioSNAP Datasetes: Stanford Biomedical Network Dataset Collection}.

\end{thebibliography}

\end{document}